# Wall Jet Similarity of Impinging Planar Underexpanded Jets

Patrick Fillingham[1] and Igor V Novosselov[1,2,*]

[1]*Department of Mechanical Engineering, University of Washington, Seattle, WA 98195, USA*
[2]*Institute for Nano-Engineered System, University of Washington, Seattle, WA 98195, USA*

## ABSTRACT

Velocity profiles and wall shear stress values in the wall jet region of planar underexpanded impinging jets are parameterized based on nozzle parameters (stand-off height, jet hydraulic diameter, and nozzle pressure ratio). Computational fluid dynamics is used to calculate the velocity fields of impinging jets with height-to-diameter ratios in the range of 15 to 30 and nozzle pressure ratio in the range of 1.2 to 3.0. The wall jet has an incomplete self-similar profile with a typical triple-layer structure as in traditional wall jets. The effects of compressibility are found to be insignificant for wall jets with $Ma<0.8$. Wall jet analysis yielded power-law relationships with source dependent coefficients describing maximum velocity, friction velocity, and wall distances for maximum and half-maximum velocities. Source dependency is determined using the conjugate gradient method. These power-law relationships can be used for the mapping of wall shear stress as a function of nozzle parameters.

## I. INTRODUCTION

Impinging jets have been studied extensively; their characterization is used in many engineering applications. Most studies focus on heat and mass transfer [1-4], this work analyzes the properties of the wall jet originating from the impingement of underexpanded planar jets, with applications to surface cleaning and non-contact micro-particle sampling. Previous studies of underexpanded jets have generally been motivated by the flow dynamics and acoustics of a short takeoff and vertical landing aircraft [5-7]; however, the wall jet region of these systems has not been studied extensively. Additionally, previous studies did not address the wall jet for planar (high aspect ratio rectangular) geometries. In some respects, the analysis of these planar jets is less complicated as 2D approximation can be used. The main advantages of underexpanded planar jets are: (i) when compared to axisymmetric jets, the planar jets are able to sustain greater wall jet velocity, thus producing higher wall shear stress further from impingement point, and (ii) planar jets cover a larger area for cooling application, particle sampling, and surface cleaning. Unerexpanded jets are studied as they provide high velocity in the wall jet even for large standoff distances. Underexpanded jets also allow for the use of isentropic nozzle relations in calculations of fluid properties at the exit of underexpanded jets, which is convenient for use in numerical simulations as boundary or initial conditions and interpretation of experimental data.

In applications related to aerodynamic particle resuspension, it is useful to characterize the wall shear stress originating from jet impingement. Measuring wall shear stress is challenging. Young et al. [8] used oil-film interferometry to measure the shear stress from an impinging supersonic jet. Their experiment has shown promise, but oil-film interferometry is limited in its precision. Tu & Wood [9] studied the wall shear stress developed from subsonic impinging jets using Preston and Stanton tube measurements; however, their results were affected by the measurement apparatus, and their conclusions are difficult to extrapolate to compressible jets. Smedley et al. [10] and Phares et al. [11] investigated the removal of microspheres from impinging jets and used theoretical relationships to infer the wall shear stress on the plate. Shear stress was found to be related to particle forces, but the study did not account for compressibility and turbulent effects. Relating particle adhesion and drag forces in the aerodynamic particle removal scenario requires multiple assumptions related to particle and surface properties; velocity measurements near the wall can be used to directly evaluate the shear stress. Loureiro demonstrated the use of laser doppler anemometry (LDA) for measuring velocity within 50 micrometers of the wall; however, the viscous layer thickness relevant to microparticle removal can be significantly smaller ~ 20µm [12]. Keedy et al. [13], using Birch's [14] model for underexpanded jets, also illustrated that organic particles could only be removed with high-pressure, axisymmetric jets at low standoff distances; however shear stress characterization was proven difficult. There is a scarcity of reliable experimental wall shear stress data in the scientific literature, especially for compressible and planar impinging jets, which suggests that numerical and analytical modeling are needed to provide insights into the wall jet behavior.

---

[*] ivn@uw.edu

The planar wall jet has been studied; most studies are based on a flow developing from the point of attachment to a wall and do not account for flow impingement and the relationship to the momentum source. Thus, it is unclear if the previous results related to wall jet similarity formulations would hold for the impinging jet scenario. The wall jet resulting from axisymmetric impinging jets has been studied experimentally [15, 16]; the studies show that the wall jet developed downstream of impingement does demonstrate self-similar behavior.

We present a parametric study that characterizes the velocity profile and wall shear stress of the wall jet resulting from impingement of planar underexpanded jets. The velocity fields from a parametric study are examined to provide a parametric mapping of the velocity and wall shear stress in the wall jet region of the flow. Wall jet velocity profiles from the CFD are presented in similarity coordinates, and the similarity variables of maximum velocity, friction velocity, maximum velocity location, and half maximum velocity location are calculated for each case of 25 cases at 20 x-locations. These variables are then normalized and calculated as a function of x-location and nozzle parameters (stand-off height, jet hydraulic diameter, and nozzle pressure ratio) for each case in the form of power-laws with source dependent coefficients. The exponents of the power-laws and the exponents for the dependence of the coefficients on the nozzle parameters are found via a least-squares regression using the conjugate gradient method to minimize the error across all computational cases.

### A. Wall Jet Theory

For mapping of flow properties near the surface and wall shear stress, it is useful to examine the wall jet portion of the flow from a similarity perspective. FIG. 1 shows the schematic diagram of the impinging jet system. Similarity variables are obtained by normalizing by x-dependent variables; $y_m$ - wall-normal location of maximum velocity, $y_{1/2}$ - wall-normal location of half-maximum velocity in the outer layer, $u_m$ - maximum wall jet velocity, and $u_\tau$ - wall jet friction velocity. The planar turbulent wall jet has consistently been shown to have incomplete similarity, which is to say that a non-dimensional similarity solution cannot describe the velocity profile of the wall jet without Reynolds number or scale dependence. Thus, one must separate the wall jet into three regions: a self-similar wall layer where viscous forces are dominant, a self-similar outer layer which behaves analogously to a free jet, and an overlap layer with source dependence where the velocity is closest to the maximum. A triple-layered incomplete similarity is achieved by matching the self-similar outer and wall regions with the overlap layer. Source dependence has been studied for true wall jets but is not defined for the wall jets resulting from impinging jets.

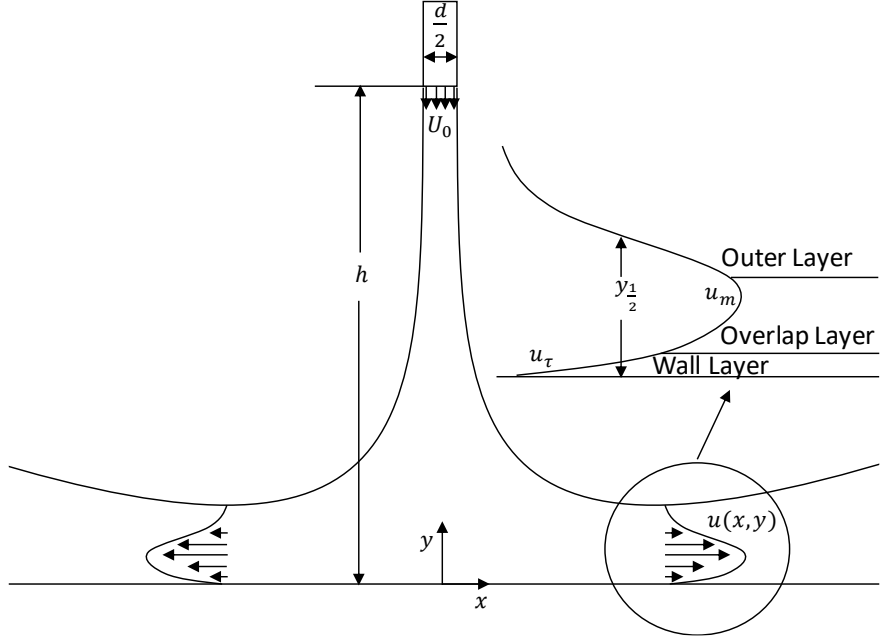

FIG. 1. Schematic of an impinging jet and the resulting wall jet. $h$ - standoff height, $d$ - jet hydraulic diameter, which is equal to twice the jet width for 2D planar jets, $y_{1/2}$ - location of half maximum velocity, $u_m$ - maximum wall jet velocity, and $u_\tau$ - wall jet friction velocity.

The equation of motion for the wall jet is defined as:

$$u\frac{\partial u}{\partial x} + v\frac{\partial u}{\partial y} = \frac{\partial}{\partial y}\left[\nu\frac{\partial U}{\partial y} - \frac{\tau}{\rho}\right] \qquad (1)$$

$$u \to 0 \text{ as } y \to \infty; \quad u = v = 0 \text{ at } y = 0.$$

As first proposed by Glauert [17], the equations of motion are assumed to be solved by outer and inner self-similar equations. The outer region becomes:

$$u = u_m(x)f_o'(\eta) \qquad (2)$$
$$\eta = \frac{y}{y_{1/2}(x)}.$$

George et al. [18] demonstrated that the classical "law of the wall" coordinates for turbulent boundary layers can be used for turbulent wall jets:

$$u = u_\tau(x)f_i(y^+) \qquad (3)$$
$$y^+ = \frac{yu_\tau(x)}{\nu}.$$

The properties of inner and outer regions need to be merged in what has traditionally been called the overlap region. George et al. [18] concluded that the overlap velocity profile could be accurately described in both inner and outer similarity coordinates, but Gertsen [19] demonstrated that the velocity in this overlap region could be more accurately described in the form of a defect law:

$$u = u_m(x) - u_\tau(x)f'(\eta_m) \qquad (4)$$
$$\eta_m = \frac{y}{y_m(x)}.$$

The solutions to these similarity equations have been determined separately by George [18] and Gertsen [19]. In this work, we examine the x-dependent variables, which can be used to describe the rest of the flow field when determined. Thus, we are interested in developing relations for $y_{1/2}$, $y_m$, $u_m$, and $u_\tau$. For each of these variables, we will assume a power-law relation in $x$ [20] with source dependent coefficients:

$$y_{1/2} \sim \beta_1 x^{\alpha_1}, \quad y_m \sim \beta_2 x^{\alpha_2}, \quad u_m \sim y_{1/2}^{\alpha_3}, \quad u_\tau \sim \beta_4 x^{\alpha_4}.$$

To determine the power-law exponents, one must determine proper scaling through dimensional analysis. In the description of the planar impinging jet, we consider seven parameters: $x \sim L$, the streamwise distance from the impingement point; $y \sim L$, the distance from the impingement surface; $h \sim L$, the standoff height of the jet; $d \sim L$, the jet hydraulic diameter; $\rho \sim ML^{-3}$, the fluid density; $\nu \sim L^2 T^{-1}$, the kinematic viscosity; and $U_0 \sim L^1 T^{-1}$, the velocity at the jet exit. L, M, and T are the units of length, mass, and time, respectively. Using these variables for dimensional analysis yields the following non-dimensional groups:

$$\Pi_1 = \frac{h}{d}, \Pi_2 = \frac{x}{h}, \Pi_3 = \frac{y}{h}, \Pi_4 = \frac{U_0 d}{\nu}.$$

Narashima et al. [21] demonstrated that scaling x and y by the momentum flux of the source is effective when writing power-laws for the velocity in wall jets, while George et al. [18] defines the momentum flux as $M_o = U_0^2 d/2$. In the study of underexpanded jets, one must consider the changes in density by defining the momentum flux as $J = \rho_0 U_0^2 d/2$. This normalization yields the following non-dimensional versions of $x, y, u_\tau,$ and $u_m$:

$$X = \frac{Jx}{\rho_\infty \mu_\infty}, \quad Y_{1/2} = \frac{Jy_{1/2}}{\rho_\infty \mu_\infty}, \quad Y_m = \frac{Jy_m}{\rho_\infty \mu_\infty}, \quad U_\tau = \frac{u_\tau \mu_\infty}{J}, \quad U_m = \frac{u_m \mu_\infty}{J}.$$

It is important to note that this procedure does not account for all of the source dependence. To account for the incomplete self-similarity of wall jets, one must consider a Reynolds number associated with the jet width. To capture the physics of underexpanded jets, the nozzle pressure ratio (NPR) and standoff height to jet hydraulic diameter ratio considered:

$$Re_n = \frac{U_0 d}{\nu_\infty}, \quad NPR = \frac{P_0}{P_\infty}.$$

Wygnanski et al. [22] established that $Y_{1/2}, Y_m, U_\tau,$ and $U_m$ can be expressed as power-laws of the form:

$$Y_{1/2} = \beta_1 X^{\alpha_1} \quad (5)$$

$$Y_m = \beta_2 X^{\alpha_2} \quad (6)$$

$$U_m = \beta_3 Y_{1/2}^{\alpha_3} \quad (7)$$

$$U_\tau = \beta_4 X^{\alpha_4}. \quad (8)$$

For the case of underexpanded impinging jets, we can assume each beta term is a function of the nozzle parameters $Re_n$, $h/d$, and $NPR$ of the form $\beta = Re_n^a (h/d)^b NPR^c$. The alpha terms are assumed to be universal across all cases. By assuming solutions for the wall jet variables of this form, we can determine the exponents in the coefficients and the power-law exponents simply by linear least squares regression on the natural logarithm of equations 5-8, i.e.: $\ln(Y_{1/2}) = a \ln(Re_n) + b \ln(h/d) + c \ln(NPR) + \alpha_1 \ln(X)$.

### B. Compressibility Effect in the Wall Jet Region

While underexpanded impinging jets provide high wall shear stress, which is desirable for aerodynamic particle resuspension, flow in the wall jet region is compressible and has the potential to introduce complications in similarity formulations. The effects of density fluctuations on turbulence have been shown by Morkovin [23] to be negligible for compressible jets for $Ma < 1.5$. The range of cases in this work is limited to subsonic wall jets ($Ma < 0.8$), so the turbulent properties are not likely to be affected by compressibility. However, mean density effects may still be important. Ahlman et al. [24] found, through a direct numerical simulation (DNS) study, that mean density effects were only significant in the wall-normal direction by comparing Reynolds and Favre averaged velocity profiles for the outer layer and comparing traditional wall coordinates with semi-local [25] and Van Driest [26] scaling. When examining velocity profiles, it was also found that mean density effects were minimal. Plotting profiles in Van Driest and semi-local scaling did not yield a noticeable improvement in similarity analysis, as shown in SI figures 1-3. For this reason, the effects of compressibility on wall jet similarity are not considered for the range of Mach numbers presented in this work. It is likely that this assumption is not valid for transonic and supersonic wall jets.

## II. COMPUTATIONAL STUDY

### A. Computational Method

The scientific literature does not report the experimental data or DNS related the wall jet developed from compressible impinging jets in the literature. To compute the flow properties needed for estimation of shear stresses, we use steady-state CFD simulation. While Shukla and Dewan [27] found that LES and DES can be accurate in predicting flow profiles, these methods are computationally intensive for a broad parametric study. Numerical simulations for this work were performed using ANSYS FLUENT 17.2 software and a $k - \omega$ shear stress transport closure model. The pressure-velocity coupled algorithm known as the QUICK (Quadratic Upstream Interpolation for Convective Kinematics) scheme [28] was used to solve the steady-state Favre-Averaged Navier-Stokes equations:

$$\frac{\partial(\bar{\rho}\widetilde{u_i})}{\partial x_i} = 0 \tag{9}$$

$$\frac{\partial(\bar{\rho}\widetilde{u_i}\widetilde{u_j})}{\partial x_i} = -\frac{\partial \bar{p}}{\partial x_i} + \frac{\partial \overline{\tau_{ij}}}{\partial x_j} - \frac{\partial\left(\overline{\rho u_i'' u_j''}\right)}{\partial x_j} \tag{10}$$

$$\frac{\partial}{\partial x_j}\left(\bar{\rho}\widetilde{u_j}\left(\tilde{h} + \frac{1}{2}\widetilde{u_i}\widetilde{u_i}\right) + \widetilde{u_j}\overline{\rho u_i'' u_i''}\right) = \frac{\partial}{\partial x_j}\left(\widetilde{u_i}(\overline{\tau_{ij}} - \overline{\rho u_i'' u_j''}) - \bar{q} - \overline{\rho u_j'' h''} + \overline{\tau_{ij} u_i''} - \frac{1}{2}\overline{\rho u_j'' u_i'' u_i''}\right). \tag{11}$$

While turbulence closure models are known to be flawed, especially when modeling impinging jets, Jaramillo et al. [29] demonstrated that $k - \omega$ models can accurately calculate the mean flow of planar impinging jets when compared to DNS. The $k - \omega$ shear stress transport (SST) model used in this work, uses $k - \epsilon$ away from the wall in the free stream and free jet portions of the flow while using $k - \omega$ near the wall to resolve the boundary layer. As demonstrated by Alvi et al. [30] and discussed by Fillingham et al. [31], the $k - \omega$ SST model [32] is a good choice for modeling underexpanded impinging jets while resolving the wall jet boundary layer. Shukla and Dewan. [33] also found $k - \omega$ SST to be superior to other closure models when considering planar impinging jets.

Figure 2 shows the computational domain. The inlet boundary condition is defined as the exit of an isentropic nozzle where the flow is choked; thus, the boundary can be described by a total pressure and a static pressure where the total pressure is necessarily (for an ideal diatomic gas) 1.893 times the static pressure. The walls are modeled as isothermal, no-slip boundaries. The outlets are defined as atmospheric pressure outlets. The outlets are located at 50 jet hydraulic diameters (100 jet slot widths) from the jet axis corresponding to a minimum of 1.5 times the impingement height.

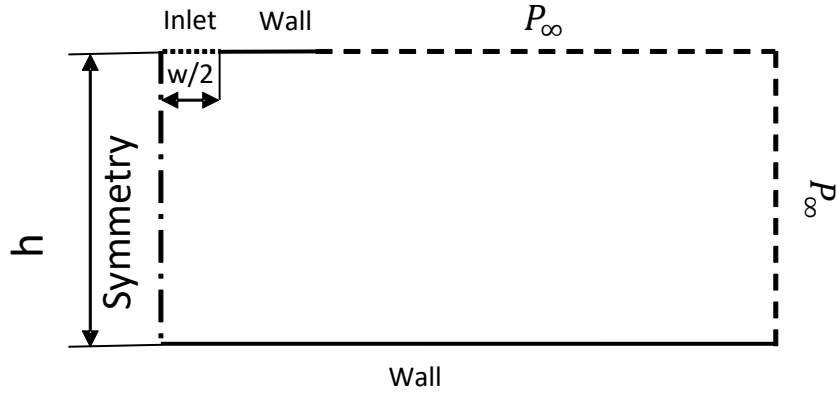

FIG. 2. Schematic of CFD domain and boundary conditions. Note that $w = d/2$ for an infinite planar jet, where d is the hydraulic diameter.

The computational grid contains ~600,000 quadrilateral elements. At the impingement surface, the first node is at a constant distance from the wall and is placed within a $y^+$ value of 1 at the maximum shear stress location, ensuring that the viscous sublayer is resolved for the entirety of the domain. The x-direction spacing is set to avoid the elements with an aspect ratio greater than 50:1. Mesh independence was confirmed by doubling the number of elements; this further mesh refinement did not affect the results, see SI figure 4. Table I shows the conditions used in the study. The range is chosen based on the wall shear stress required for microparticle resuspension [12, 13]. All cases result in subsonic wall jet. For a supersonic wall jet, a separate characterization would be necessary; these are beyond the scope of the paper.

TABLE I. Summary of CFD Cases.

| h/d | d (mm) | NPR |
|---|---|---|
| 15, 17.5, 25, 30 | 1, 2 | 1.2, 1.6, 2.0, 2.4, 2.8 |

### B. Evaluation of the CFD Approach

To evaluate the 2D assumption, a 3D simulation of the jet with an aspect ratio of 30 to 1 was performed. As with the 2D simulation, QUICK scheme and $k - \omega$ SST turbulence closure model were used. The 3D domain includes ~12 million elements with the first node in the wall-normal direction at $y^+ = 1$. FIG. 3 shows that the centerline profile of wall shear stress from the 3D simulation agrees well with the 2D simulations.

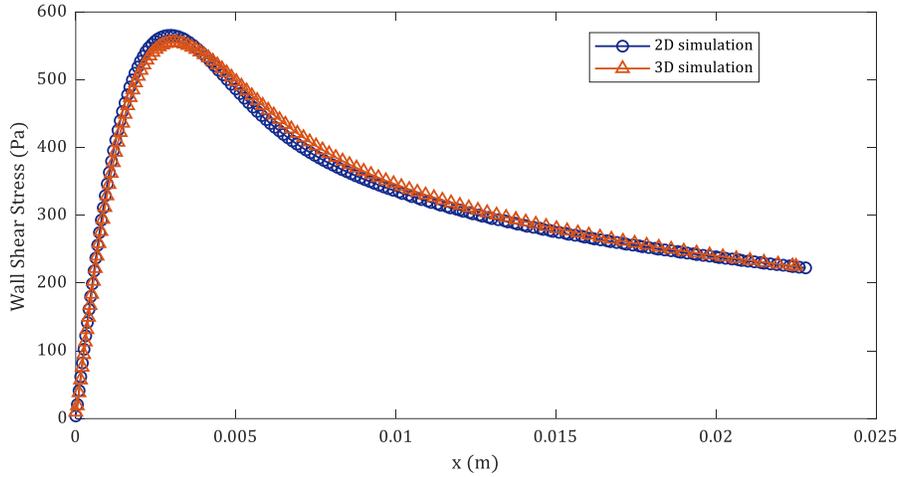

FIG. 3. Comparison of wall shear stress as a function of distance from impingement location for 2D and 3D simulations: h = 30 mm, d = 1 mm, and NPR = 2.0 from the 2D simulation and centerline profile from the 3D simulation.

Validation of the CFD result is challenging in the absence of the experimental or DNS data for compressible planar impinging jets needed for direct comparison. To validate the model, we evaluated two flow regions: (i) impinging jet and (ii) wall jet region. Related to the impinging jet region, implementation of $k - \omega$ SST jet was previously validated in the study of underexpanded axisymmetric jets. The shape and the shock structures of the impinging jet predicted by the $k - \omega$ SST model were shown to be in excellent agreement with experimental observations from Schlieren photography [31]. This provides confidence in the modeling accuracy of the supersonic region of the jet. Additionally, in this study, the CFD predictions were compared with normal pressure profiles on the impingement surface from pressure-sensitive paint (PSP) experiments. PSP utilizes the emission spectra of a luminophore by relating the emission intensity at specific wavelengths to the partial pressure of oxygen at the surface. Images were taken at wind-on and wind-off conditions, and the intensity ratio of the images relate to pressure. Binary FIB PSP and a PSP-CCD camera from Innovative Scientific Solutions Incorporated (ISSI) (Dayton, OH, USA) were used. The calibration curve measurements for the PSP were provided by ISSI as well. Evaluation of the numerical model was performed using oblique planar jet impingement against the PSP measurements. Oblique impingement results in an uphill shift in the impingement point from the geometric center [34]. Figure 4 shows the CFD pressure profiles overlayed on the PSP measurements. The CFD simulations show agreement in shape and magnitude with PSP measurements of the pressure profile, allowing for confidence in the accuracy of the CFD in the impingement region.

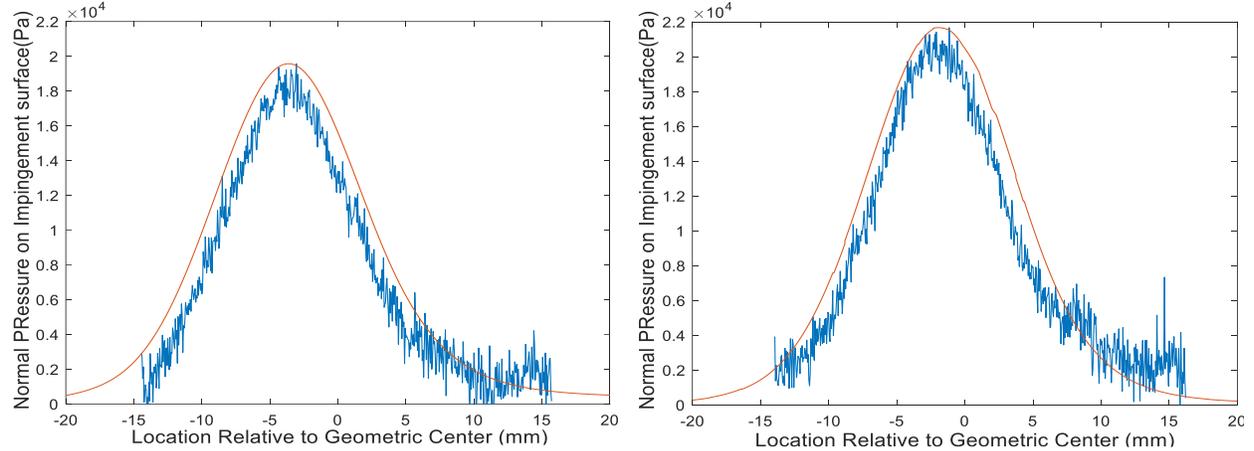

FIG. 4. Normal pressure profiles from CFD (red) and pressure-sensitive paint (blue) experiments for h = 30 mm, d = 1 mm, NPR = 1.0 with the impingement angle of 30 degrees (left) and 15 degrees (right).

To evaluate the model accuracy in the wall jet region, the $k-\omega$ SST model results are compared to two separate DNS studies. First, the model was used to replicate DNS data from planar impinging jet conducted by Jaramillo et al. [29] The authors report a wall jet velocity profiles at the locations downstream of the impingement point, up to 8 jet widths. Figure 5 (a) shows the velocity profile from DNS by Jaramillo et al. [29] and $k-\omega$ SST are in good agreement. This agreement is closer for locations away from the impingement point, which is of the most interest in our study. Second, the model was evaluated against DNS of a classical wall jet conducted by Naqvi et al. [35] the boundary conditions were matched to ensure the wall jet development was modeled accurately. Figure 5 (b-d) compares the development of the wall jet thickness, $y_{1/2}$, the maximum velocity, $U_m$, and the wall shear stress, $\tau$. The 2D $k-\omega$ SST model shows excellent agreement with both planar impinging jet [29] and wall jet DNS [35] studies. In summary, the $k-\omega$ SST was found to be an acceptable model for use in the parametric study of planar impinging jet due to its good agreement with the PSP measurements, Schlieren photography, as well as two DNS studies describing wall jet development.

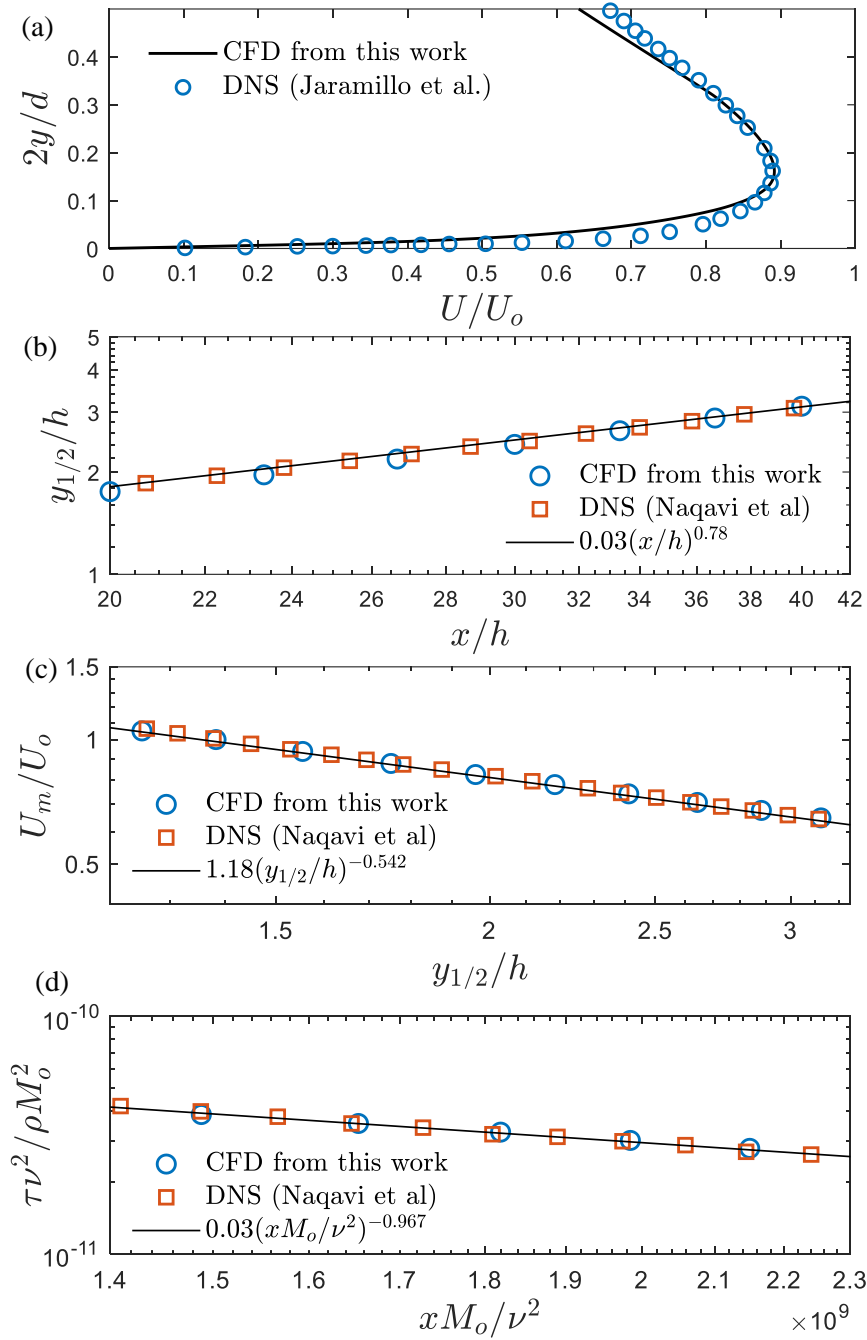

FIG. 5. Validation of $k-\omega$ SST for impinging jet modeling (a) Comparison of the velocity profile at 8 jet widths downstream of impingement location against DNS from Jaramillo et al. [29]. Comparison with wall jet DNS simulation from Naqavi et al. [35] (b) Decay of half maximum velocity location. (c) maximum velocity (d), and wall shear stress.

## III. RESULTS AND DISCUSSION

### A. Wall Jet Velocity Profiles

Velocity profiles from the 2D impinging jet simulations are examined to determine the self-similarity of the wall jet region. Traditionally, the planar wall jet has been considered self-similar in coordinates presented in Eq. 2. Wygnanski [22] observed that normalization by $y_{1/2}$ and $u_m$ appears to yield similarity for the entire velocity profile; however, it was later demonstrated [18] that normalization by $y_{1/2}$ and $u_m$ only yields similarity in the outer region (y > y$_{1/2}$) of the jet. The velocity profile for the outer region is identical to that of a free jet and thus can be described by:

$$f'_o = 1 - (\tanh k\eta)^2 \tag{12}$$

$$k = \operatorname{atanh}\sqrt{\frac{1}{2}}.$$

Figure 6 illustrates the self-similarity in the outer region for three different geometries and NPRs, comparing the CFD simulations to the analytical solution in Eq. 12. The self-similarity develops downstream of the impingement point for $x/h > 0.2$.

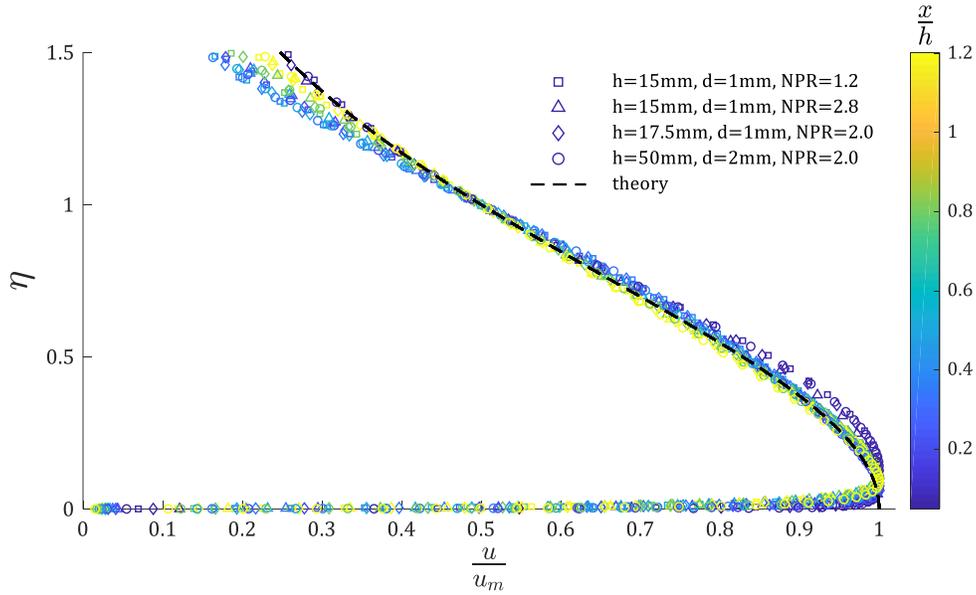

FIG. 6. Velocity profiles plotted in outer coordinates for four different cases vs. the theoretical profile (Eq. 12). Profiles demonstrate similarity independent of the geometry and nozzle pressure ratio.

To examine similarity in the overlap layer, the coordinates described by Eq. 4 are used in a defect relation given by Eq. 13. Gertsen [19] developed an analytical expression for the velocity profile:

$$f' = \frac{1}{0.41}\left(-\ln \eta_m - \frac{5}{6} + \frac{3}{2}\eta_m^2 - \frac{2}{3}\eta_m^3\right). \tag{13}$$

In Figure 7, velocity profiles are plotted in defect coordinates for two geometries and two NPRs; the simulations are also compared with the analytical expression, Eq. 13. The overlap layer similarity takes longer to develop ($x/h > 0.4$) than the inner and outer layers.

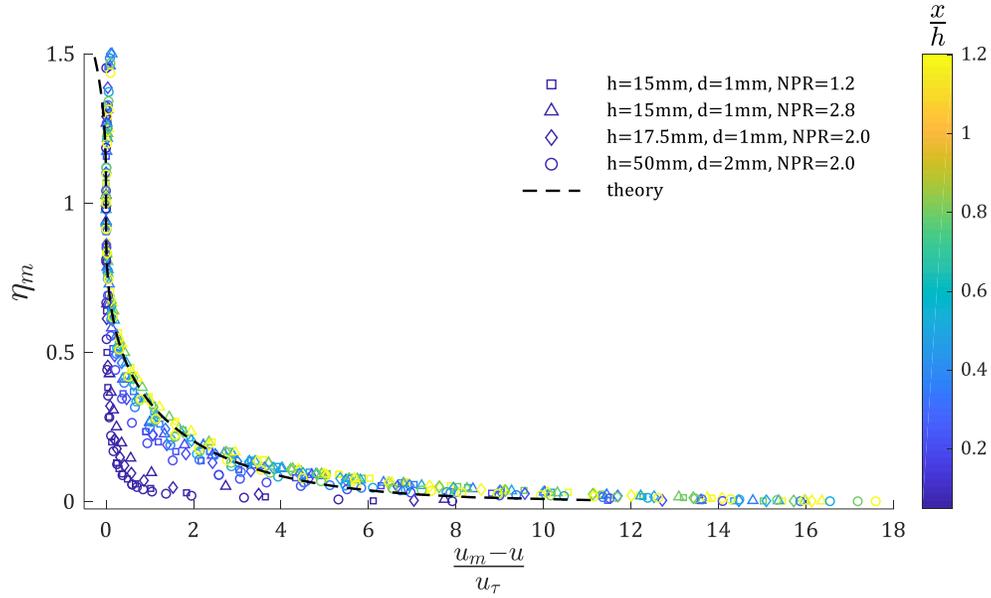

FIG. 7. Velocity profiles plotted in defect coordinates for four different cases vs. the theoretical profile (Eq.13). Profiles demonstrate similarity independent of the geometry and nozzle pressure ratio.

When examining the velocity profile in defect coordinates (FIG. 6), it appears that similarity extends to the wall layer; however, the analytical expression for the velocity profile derived from the equations of motion does not apply for $y^+ < 30$. To obtain similarity in the viscous wall layer, the velocity profiles are plotted in the traditional "law of the wall" coordinates. Figure 8 plots the wall layer for the same cases as Figures 6 and 7.

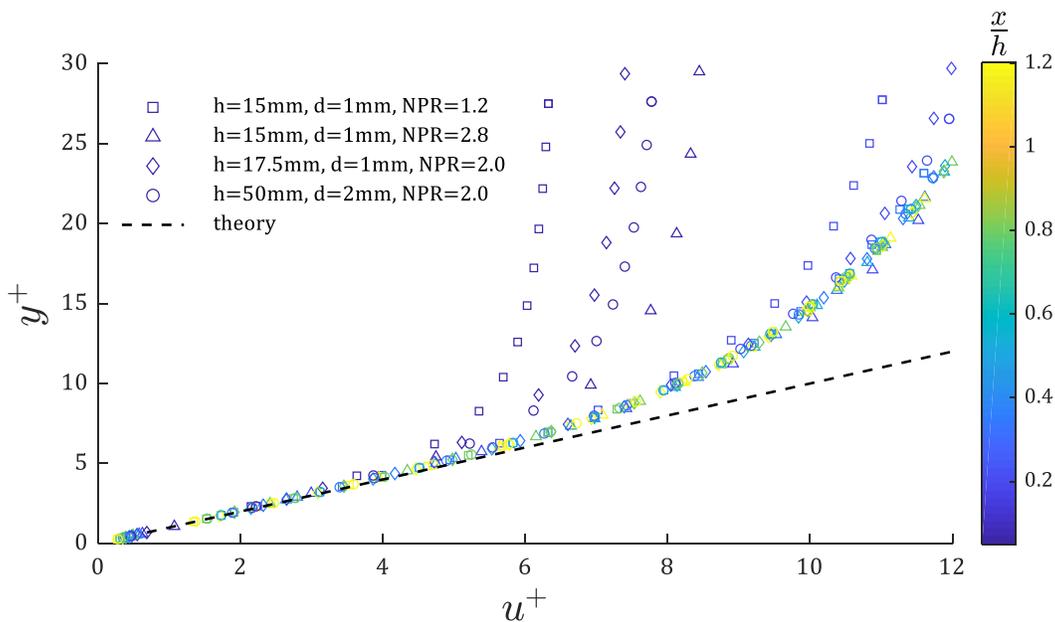

FIG. 8. Velocity profiles plotted in wall coordinates for four different cases. The $y^+ = u^+$ shown as the dash line. Profiles demonstrate similarity independent of the geometry and nozzle pressure ratio.

Figure 8 demonstrates that the inner layer of the wall jet follows the traditional law of the wall, with a linear velocity profile in the viscous sublayer up to a $y^+$ value of 5. This is important when considering particle removal as the linear velocity profile allows for the calculation of drag forces of particles in the sublayer with only knowledge of the wall shear stress. Figures 7 and 8 show that an analytical expression that characterizes the "buffer" region (between the linear and log law regions of the boundary layer) of the velocity profiles of the turbulent boundary layers does not exist. Plotting data in the established similarity coordinates shows that impinging jets produce wall jets in the same triple-layer structure demonstrated experimentally [36].

## B. Power-Law Relationships

After confirming the similarity of the wall jet velocity profile, the similarity variables from each computational case were calculated and then analyzed to obtain the source-dependent power-law relationship as a function of *x*-coordinate for each similarity variables. Before calculating the power-laws, it is important to note that while the far-field conditions are not typically considered in the wall jet analysis, they are significant when determining the power-law exponents. As Gertsen [19] pointed out, his analysis is only valid in the absence of a perpendicular wall coincident with the source of the wall jet. The analysis of George et al. [18] considers the wall jet to be emerging from a perpendicular wall and thus discrepancies between their power-law exponents are expected. An impinging jet should behave similarly to that of a wall jet emerging from a perpendicular wall. The available experimental and computational studies [37], [38] consider a wall jet emerging from a wall have all been conducted in water tanks and represent confined jet scenario for both top and side boundary conditions. In this study unconfined jets are investigated; thus we expect some variation in the power-law relationships developed in this work.

The characteristic length of wall jet velocity profiles has been debated in the scientific literature [17-19]. Generally, the distance from the wall in the outer region, where the velocity is half of the maximum ($y_{1/2}$), is used as the length scale. Though it was suggested that this choice is arbitrary, the use of $y_{1/2}$ has repeatedly [21, 22, 38] shown to be useful in characterizing the similarity of wall jet velocity profiles. It was also demonstrated that momentum normalized $y_{1/2}$ can be accurately described by a source dependent power-law in the *x*-direction with a virtual origin. Figure 9 plots $Y_{1/2}$ against $X$ for all geometries with 1 mm jet hydraulic diameter and all NPRs. Simulations show that a virtual origin is necessary for similarity, which is consistent with the previous reports [18, 22]. While there is not an obvious physical choice for the virtual origin of a traditional wall jet, the standoff height is a logical choice for impinging jets. Here, we define the virtual origin location as $X_0 = \frac{-Jh}{\rho_\infty \mu_\infty}$.

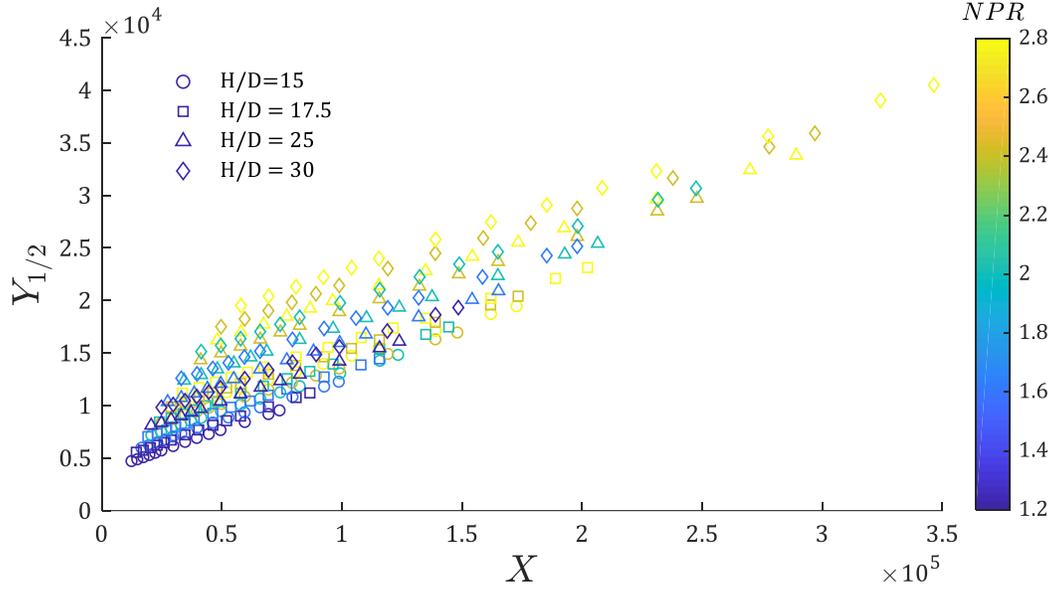

FIG. 9. Momentum normalized half velocity wall distance plotted against momentum normalized x-location for all height-to-diameter ratios colored by the nozzle pressure ratio.

Figure 10 (a) demonstrates the effectiveness of using standoff height as a virtual origin. The similarity is nearly obtained; however, an adjustment for source dependence based on nozzle pressure ratio improves the fit; $\beta_1 \sim NPR^{0.15}$ yields a linear relation, as shown in Figure 10 (b).

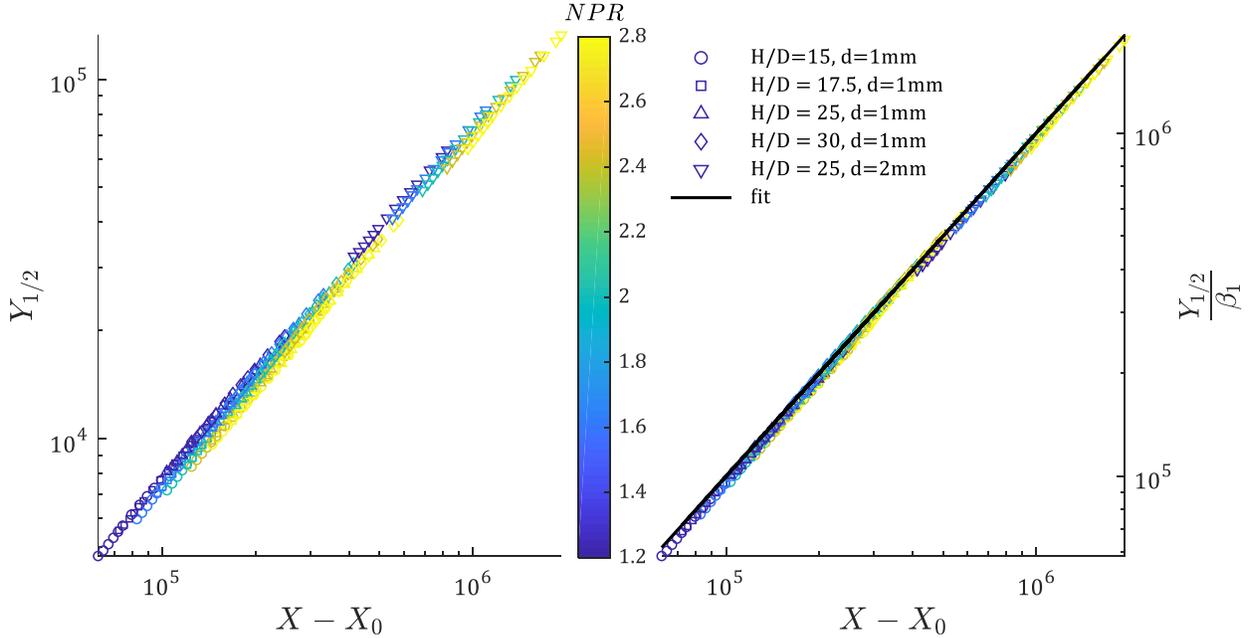

FIG. 10. Momentum normalized $Y_{1/2}$ plotted against the momentum normalized x-location with a virtual origin at the jet nozzle location and analytical solution to Eq. 14: (a) without nozzle pressure source dependence adjustment; (b) with nozzle pressure source dependent adjustment.

George et al. [18] proposed that the dependence of $y_{1/2}$ on $x$ is necessarily non-linear but approaches linearity in the limit of $Re \to \infty$; for the experimental case examined $\alpha_1 = 0.97$. Gertsen [19] suggested that a linear relationship should be expected, but this determination is dependent on the absence of a wall perpendicular to the wall jet source. Naqavi [35] found that for the case of co-flow the best fit was for $\alpha_1 = 0.72$; however, the maximum x-coordinate in the analysis was limited to forty jet widths, and the wall jet was not fully developed. Since the virtual origin was not considered and the fit is weighted to the thinner boundary layer the exponent is significantly different from the analytical expressions for fully developed flow [18], [19]. In our analysis the wall jet is not constrained thus fully developed flow can be achieved; the least-squares analysis in this yielded $\alpha_1 = 0.98$.

$$Y_{1/2} = \beta_1 (X - X_0)^{0.98} \quad (14)$$

$$\beta_1 = 0.083 * NPR^{0.15}$$

In characterizing maximum velocity in the wall jet, a power-law based on a local length scale can be more accurate than one based on the global $x$-coordinate [18]. Intuitively, $y_m(x)$ can be used as the length scale for characterizing $u_m$. However, $y_{1/2}(x)$ has shown to have better correlation; it is also easier to measure experimentally [38] and with DNS [35]. George et al. [18] proposed that the decay exponent for $u_m$ as a function of $y_{1/2}$ is universal for wall jets. Figure 11 plots momentum normalized maximum velocity against $y_{1/2}$ with and without pressure source adjustment. Our calculations show that the NPR is the only source adjustment needed to obtain similarity in the wall jet, which is consistent with the findings that a power-law for maximum velocity based on the local length scale is universal [18].

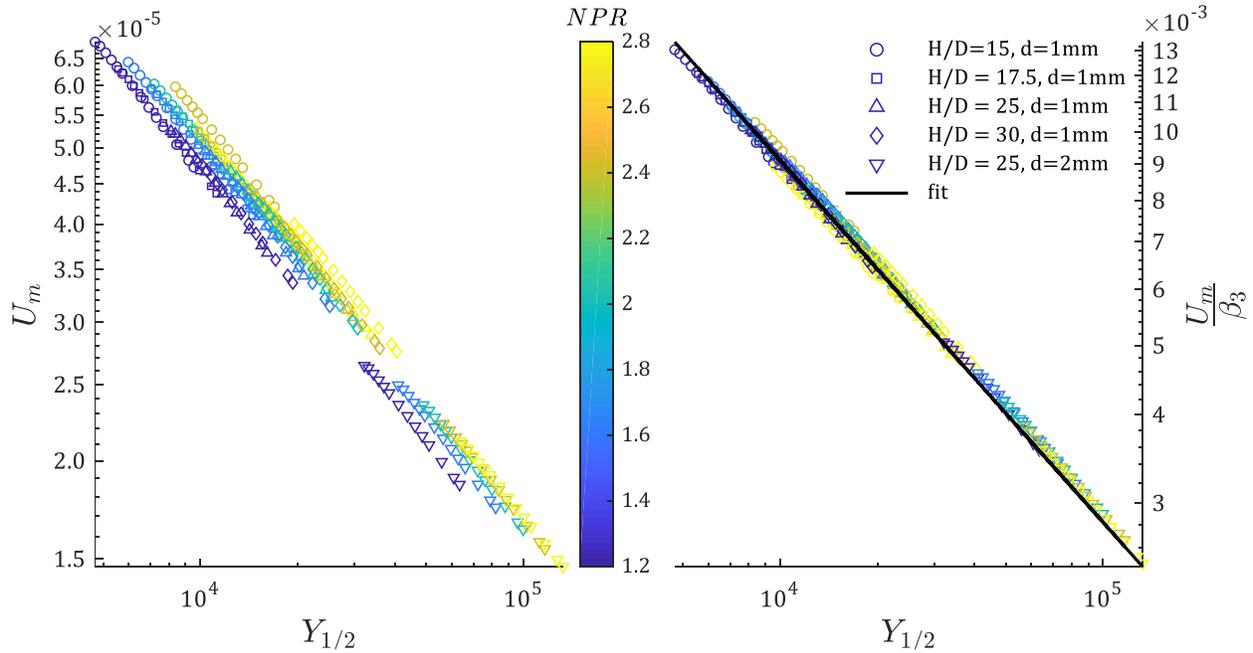

FIG. 11. Momentum normalized maximum velocity plotted against momentum normalized half-maximum velocity location: (a) without nozzle pressure source dependence adjustment; (b) with nozzle pressure source dependent adjustment and fit (Eq. 15).

After adjusting for nozzle pressure ratio, the power-law for maximum velocity relation is:

$$U_m = \beta_3 Y_{1/2}^{-0.52} \quad (15)$$

$$\beta_3 = 0.0051 * NPR^{0.15}.$$

The exponent for the decay of maximum velocity exponent $\alpha_3 = -0.52$, which is in good agreement with the analytical expression $\alpha_3 = 0.527$ [18]. The relationship between the local length scale and maximum velocity is one of the main characteristics of the traditional wall jets. The consistency with the literature provides evidence that wall jets developed from impinging jets exhibit the same length scale dependence as simple wall jets, regardless of far-field conditions.

To use defect law coordinates, the maximum velocity location, $y_m$ as a function of $x$, must be characterized. The similarity of wall jets generally assumes the ratio, $\gamma = y_m/y_{1/2}$, to be constant, however, it is only strictly true as $x \to \infty$ [18, 19]. For impinging jets near the impingement point, this approach is not valid, thus a separate power-law for $y_m$ is required. Figure 12 (a) plots momentum normalized maximum velocity location against momentum normalized $x$, showing that a virtual origin is not necessary. The source dependence, as determined by a least-squares fit, is plotted in Figure 10 (b). The final expression for $y_m$ is:

$$Y_m = \beta_2 X^{0.49} \tag{16}$$

$$\beta_2 = 0.00027 * NPR^{0.33} * \frac{h^{0.48}}{d} * Re_n^{0.85}.$$

The exponent for $Y_m$ in this work, $\alpha_2 = 0.49$, is lower than reported in the literature. Tang et al. [39] found the $\alpha_2 = 0.717$ using LDA while Naqavi et al. [35] found the $\alpha_2 = 0.743$ based on the DNS calculations. Tang et al. [39] conducted their experiments in an enclosed water tank while Naqavi et al. [35] used a coflow for the DNS. These far-field conditions significantly change the entrainment pattern. The jet would primarily entrain momentum from co-flow or large-scale structures moving in the x-direction, however, in the case of the perpendicular wall or impinging jet, the momentum must be entrained from the y-direction. This major difference in the entrainment pattern changes the wall development; the y-direction momentum entrainment hinders the wall jet spreading, thus reducing the location of the maximum velocity. To confirm this assumption, maximum velocity location data were taken from the velocity profiles provided by Shukla and Dewan [27]. These data were fitted with a power-law with the same exponent as found in this work with excellent agreement as shown in SI figure 5.

Fluid compressibility did not have an appreciable effect on the decay analysis. Fitted cases for only $Ma \sim 0.3$ and $Ma \sim 0.8$ had exponents of $\alpha_3 = -0.50$ and $\alpha_3 = -0.48$, respectively, which is a small difference compared to the exponent found through DNS simulations of a wall jet reported in the literature [35].

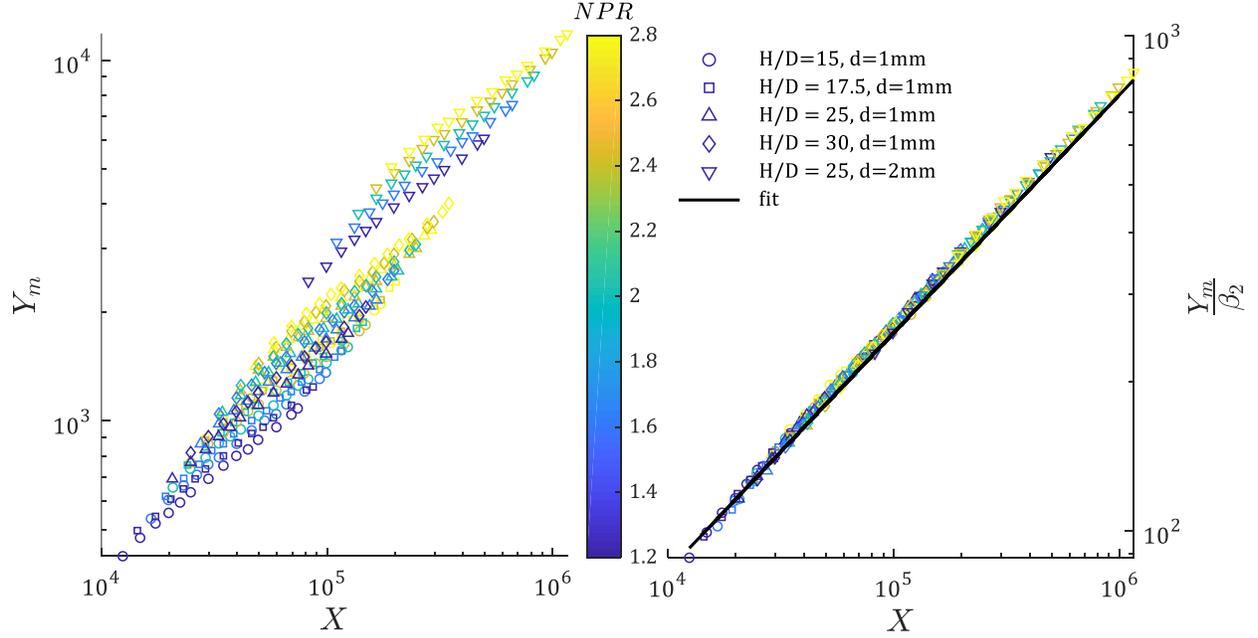

FIG. 12. Momentum normalized maximum velocity wall-distance plotted against momentum normalized x-location: (a) without nozzle pressure source dependence adjustment; (b) with nozzle pressure source dependent adjustment and fit (Eq. 16).

Friction laws are generally expressed as a friction coefficient, which is a function of a local Reynolds number. For this work, the friction coefficient power-law works for $x/h > 1.0$. Using the downstream data, a friction law has the best fit:

$$c_f = \left(\frac{u_\tau}{u_m}\right)^2 = 0.0029 Re_l^{-0.19} \tag{17}$$

$$Re_l = \frac{u_m y_{1/2}}{\nu_{0.5}}.$$

This formulation agrees with the existing literature [18, 38, 40]. Friction laws in this form are inconsistent across experimental and DNS data, however, and are highly dependent on the momentum source; thus, for this work, friction velocity was characterized directly, similar to the maximum wall jet velocity. Figure 13 demonstrates the effect of the source term adjustment. Momentum normalized friction velocity can be expressed as:

$$U_\tau = \beta_4 X^{-0.3} \tag{18}$$

$$\beta_4 = 0.021 * \frac{h^{0.22}}{d} * NPR^{-0.07} * Re_n^{-0.5}.$$

Note that it is difficult to obtain physical interpretations from the source dependent exponents, as there are insufficient analytical or experimental data of planar impinging jets. Further investigation is needed to gain insight into the source dependent exponents.

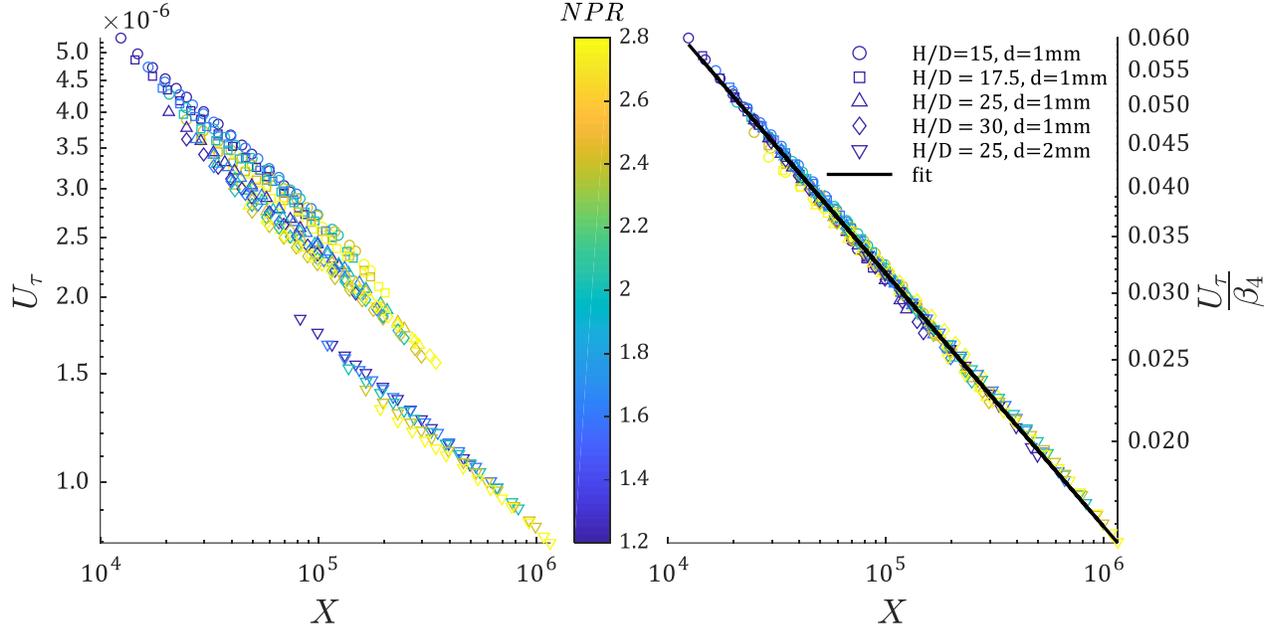

FIG. 13. Momentum normalized friction velocity plotted against momentum normalized x-location: (a) without nozzle pressure source dependence adjustment; (b) with nozzle pressure source dependent adjustment and fit (Eq.18).

## C. Wall Shear Stress

Though the fluid compressibility does not have a significant effect on the power-laws or similarity, the change in mean density is not negligible. For this reason, the wall shear stress cannot be characterized directly from friction velocity. Here, we formulate a power-law for momentum normalized wall shear stress, $\tau^*$:

$$\tau^* = \frac{\tau}{\rho_\infty}\left(\frac{\mu_\infty}{J}\right)^2.$$

Figure 14 plots momentum normalized wall shear stress against momentum normalized x with and without source dependence, demonstrating that a power-law is appropriate for wall shear stress.

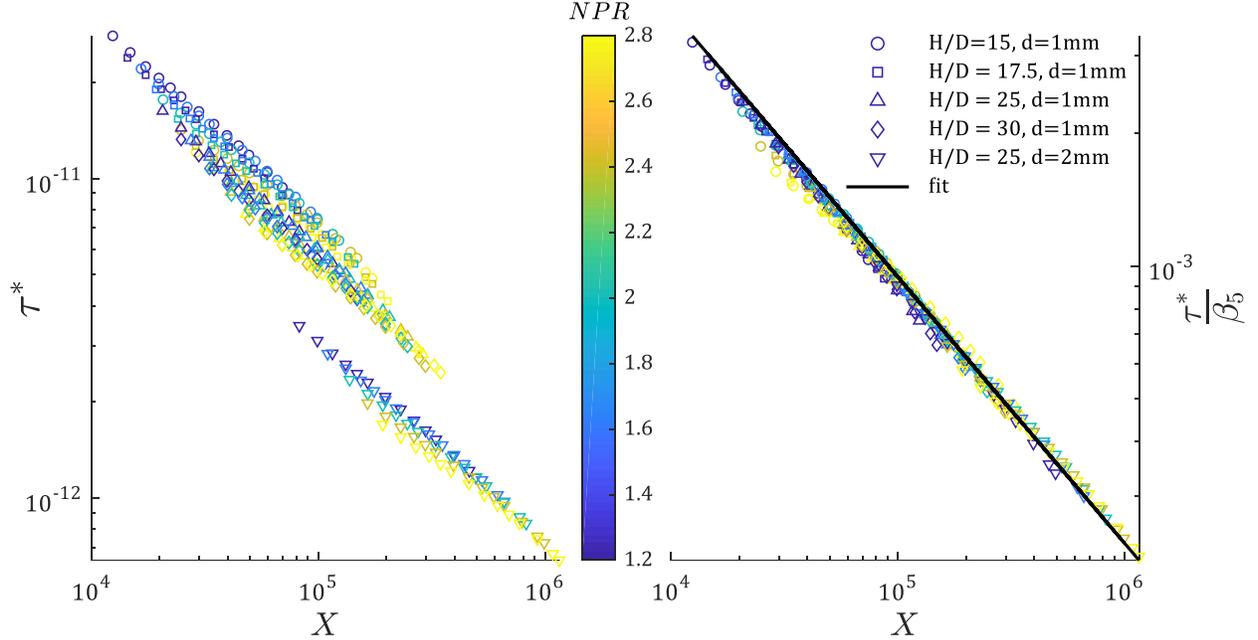

FIG. 14. Momentum normalized wall shear stress plotted against momentum normalized x-location: (a) without nozzle pressure source dependence adjustment; (b) with nozzle pressure source dependent adjustment and fit (Eq. 14).

The source dependent power-law developed for momentum normalized wall shear stress is:

$$\tau^* = \beta_5 X^{-0.61} \tag{19}$$

$$\beta_5 = 0.00059 * \frac{h}{d}^{-0.45} * NPR^{-0.18} * Re_n^{-1.0}.$$

The power-law developed in this work suggests a slower decay of wall shear stress ($\alpha = 0.61$) than those in the literature for traditional wall jets. Wygnanski et al. [22] found the decay exponent to be $-1.07$, while Naqvi et al. [35] found an exponent of $-0.967$ via DNS. As stated before, both of these studies are conducted without a wall coincident and perpendicular with the wall jet source. For a direct comparison to impinging jets, the shear stress data from Tu and Wood [9] for an impinging jet with $h/d = 20.6$ and $Re_n = 6300$ is more appropriate. The experimental data was fitted with a power-law, the exponent based on the experimental data was found to be excellent agreement as shown in Figure 15.

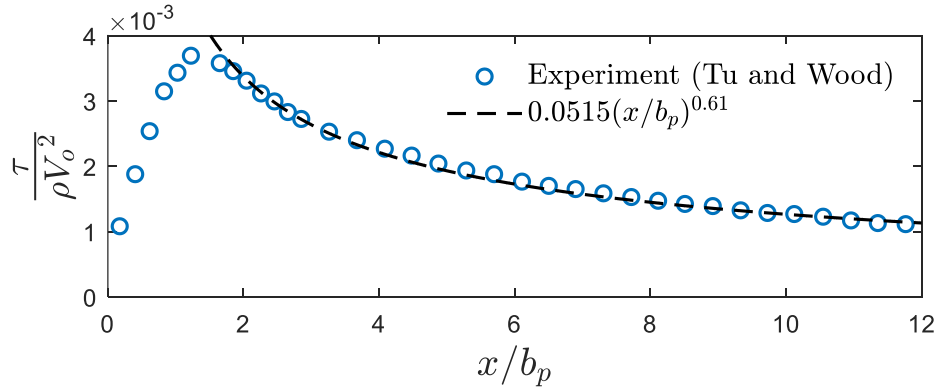

FIG. 15. Wall shear stress profile from Tu and Wood with height to jet width ratio of 20.6 and Re=6300 with power-law fit, $b_p$ is the half-width of the normal pressure profile.

### D. Summary of the Wall Jet Relations

After developing power-laws for the wall jet variables it is clear that the power-law exponents are source independent and universal while the coefficients depend on the jet nozzle parameters. This allows for the mapping of the complete wall jet velocity profile and wall shear stress with only knowledge of nozzle parameters and x location. Table II summarizes the power laws below.

TABLE II. Summary of Wall Jet Power-Laws

| Wall Jet Variable | Power-Law | Coefficient, $\beta$ |
|---|---|---|
| $Y_{1/2}$ | $\beta_1(X-X_0)^{0.98}$ | $0.083 * NPR^{0.15}$ |
| $Y_m$ | $\beta_2 X^{0.49}$ | $0.00027 * NPR^{0.33} * h/d^{0.48} * Re_n^{0.85}$ |
| $U_m$ | $\beta_3 Y_{1/2}^{-0.52}$ | $0.0051 * NPR^{0.15}$ |
| $U_\tau$ | $\beta_4 X^{-0.3}$ | $0.021 * h/d^{0.22} * NPR^{-0.07} * Re_n^{-0.5}$ |
| $\tau^*$ | $\beta_5 X^{-0.61}$ | $0.00059 * h/d^{-0.45} * NPR^{-0.18} * Re_n^{-1.0}$ |

### IV. CONCLUSIONS

A parametric study uses 2D numerical simulations to examine underexpanded impinging jets over a range of jet parameters, such as jet standoff distance, jet hydraulic diameter, and jet nozzle pressure ratio. The velocity fields calculated from CFD were transformed into similarity coordinates traditionally used for studying wall jets, demonstrating the self-similar nature of the wall jet resulting from underexpanded impinging jets. These similarity coordinates provide a framework to map the entire wall jet velocity field by developing power-law equations for the wall jet similarity variables based solely on the nozzle parameters and streamwise location. The analysis of the similarity profiles and power-laws led to the following conclusions:

- The wall jet developed from planar jet impingement has the same triple-layered structure as classical wall jets. Thus, the x-dependent length scales and velocities ($y_{1/2}, y_m, u_\tau$ and $u_m$) can be used to analyze wall jet properties.
- Compressibility does not significantly affect the similarity analysis; that is, density adjusted similarity coordinates do not yield improvement over traditional coordinates for wall jets with $Ma < 0.8$.
- Normalization by momentum, as opposed to length scales, is found to be effective in reducing source dependence in the power-laws.
- Jet geometry and operating conditions (h, d, NPR) have a significant effect on the coefficients of the power-laws, while the power-law exponents are independent of these parameters.
- The entrainment patterns have a significant effect on the power-law exponents but not the shape of the wall jet velocity profile.
- The wall jet velocity profile from under expanded impinging jets can be mapped using power-laws for $y_{1/2}, y_m, u_\tau$ and $u_m$, with only knowledge of the nozzle parameters.
- A power-law was developed for normalized wall shear stress, allowing for the prediction of wall shear stress, within a maximum error of 8%, as a function of only jet hydraulic diameter, standoff height, NPR, and x-coordinate.

## V. ACKNOWLEDGMENTS

This work was supported by the DHS Science and Technology Directorate, and UK Home Office [contract no. HSHQDC-15-C-B0033 and was facilitated using advanced computational, storage, and the networking infrastructure provided by the Hyak supercomputer system at the University of Washington.